\documentclass[a4paper,12pt]{article}
\usepackage{amsmath, amssymb}
\usepackage{multirow}
\usepackage{array}
\usepackage{graphicx}
\usepackage{amssymb}
\usepackage{stmaryrd}
\usepackage{enumitem}
\usepackage[utf8]{inputenc}
\usepackage{authblk}
\usepackage{indentfirst}
\usepackage{geometry}
\usepackage{tikz}
\usepackage{float}
\usepackage{url}
\usepackage{mathtools}
\usepackage{amsmath,amsfonts,amsthm}
\usepackage{mathrsfs} 

\newcommand{\case}[1]{\paragraph*{Case #1:}}

\newcommand{\NP}{\textbf{NP}}
\usepackage[utf8]{inputenc}
\usepackage[T1]{fontenc}
\usepackage[english]{babel}
 \usepackage{multirow}
\newtheorem{theorem}{Theorem}[section]
\newtheorem{lemma}[theorem]{Lemma}

\newtheorem{question}{Question}
\newtheorem*{answer}{Answer}
\begin{document}
\date{ }
\title{A Study of NP-Completeness and Undecidable Word Problems in Semigroups}
\author[1]{Duaa Abdullah\thanks{Corresponding Author: abdulla.d@phystech.edu}}
\author[1]{Jasem Hamoud\thanks{jasem1994hamoud@gmail.com}}
\affil[1]{Department of Discrete Mathematics, Moscow Institute of Physics and Technology}
\maketitle
\begin{abstract}
In this paper we explore fundamental concepts in computational complexity theory and the boundaries of algorithmic decidability. We examine the relationship between complexity classes \textbf{P} and \textbf{NP}, where $L \in \textbf{P}$ implies the existence of a deterministic Turing machine solving $L$ in polynomial time $O(n^k)$. Central to our investigation is polynomial reducibility. Also, we demonstrate the existence of an associative calculus $A(\mathfrak{T})$ with an algorithmically undecidable word problem, where for a Turing machine $\mathfrak{T}$ computing a non-recursive function $E(x)$, we establish that $q_1 01^x v \equiv q_0 01^i v \Leftrightarrow x \in M_i$ for $i \in \{0,1\}$, where $M_i = \{x \mid E(x) = i\}$. This connection between computational complexity and algebraic undecidability illuminates the fundamental limits of algorithmic solutions in mathematics.

\noindent\textbf{MSC Classification 2020:} 68R05, 05C85 , 68Q15, 68Q17, 68Q25.

\end{abstract}
\maketitle
\section{Introduction}
Computational complexity theory stands as one of the most profound and challenging areas in theoretical computer science, with far-reaching implications across mathematics, cryptography, and algorithm design. At its core lies the fundamental question of resource-bounded computation. What problems can be efficiently solved and what problems appear to require computational resources that grow exponentially with input size? This examination delves into two pivotal aspects of complexity theory: the classification of computational problems and the boundaries of algorithmic decidability. The dichotomy between polynomial-time solvable problems (class \textbf{P}) and those for which solutions can be verified in polynomial time (class \textbf{NP}) represents perhaps the most significant open question in computer science~\cite{Cook,Karp}. As Garey and Johnson~\cite{Garey} articulated in their seminal work, this distinction has profound implications for our understanding of efficient computation. The \textbf{P} versus \textbf{NP} problem asks whether every problem whose solution can be verified quickly can also be solved quickly - a question that remains unresolved despite decades of intensive research efforts~\cite{Fortnow2009,Aaronson}.

Central to this investigation is the concept of NP-completeness, a property that characterizes \emph{hardest} problems within the class \textbf{NP}. The Boolean satisfiability problem (SAT), first proven NP-complete by Cook~\cite{Cook}, serves as a canonical example of such problems. Through polynomial-time reductions, we can establish equivalence classes of computational difficulty, demonstrating that certain problems encapsulate the inherent complexity of an entire class~\cite{Levin}. The technique of reduction not only provides a powerful tool for classifying problems but also illuminates the structural relationships between seemingly disparate computational tasks~\cite{Arora}.

Beyond the realm of polynomial-time computation lies the domain of undecidability: problems for which no algorithmic solution exists, regardless of computational resources. The study of undecidable problems traces back to the groundbreaking work of Turing~\cite{Turing} and Church~\cite{Church}, who independently demonstrated the existence of problems that cannot be solved by any algorithm. Among these, the word problem for semigroups, as established by Post~\cite{Post} and Markov~\cite{Markov}, represents a fascinating intersection of abstract algebra and computability theory. This problem asks whether two words represent the same element in a finitely presented semigroup—a question that, in the general case, admits no algorithmic solution.

This examination explores both the classification of computational problems through the lens of NP-completeness and the fundamental limits of computation through undecidable word problems in semigroups. By investigating these complementary aspects of computational complexity, we gain deeper insights into the nature of computation itself and the boundaries that constrain algorithmic solutions to mathematical problems.

This paper is organized as follows. In Section~\ref{sec1},  we observe the important concepts to our
work, we have provided a preface through some of the
important properties we have utilized in understanding
the work; in Section~\ref{sec2}, we introduced the first problem called ``Satisfiable Problem'' in Question~\ref{question1} with solution; in Section~\ref{sec3} we provide Question~\ref{question2} for construct an associative calculus (semigroup) with an algorithmically undecidable word problem.
\section{Preliminaries}~\label{sec1}
 The class \textbf{P} consists of problems that can be solved by a deterministic Turing machine $\operatorname{DTM}$ in polynomial time, meaning the algorithm's running time is bounded by a polynomial in the input size $n$. These are considered ``efficiently'', ``efficiently solvable'', ``solvable'' problems, such as array sorting or finding shortest paths in a graph using \emph{Dijkstra's algorithm}. The class \textbf{NP} includes problems for which a solution, if it exists, can be verified by a nondeterministic algorithm Turing machine $\operatorname{NDTM}$ in polynomial time. In other words, given a ``certificate'', ``certificate'' (e.g., a candidate solution), it can be checked in polynomial time (linear in  Examples include the Hamiltonian cycle problem, the traveling salesman problem, and the Boolean satisfiability problem $\operatorname{SAT}$).\par 
The central open question in computational complexity theory is whether $\textbf{P} = \NP$. Most researchers believe $\textbf{P} \neq \NP$, as $\textbf{P} = \NP$ would imply that all problems with polynomially verifiable solutions are also polynomially solvable, with profound implications for cryptography, optimization, and other fields.\par 
Polynomial reducibility is a key concept for comparing problem complexity. A problem (or language) $L_1 \subseteq \Sigma_1^*$ is polynomially reducible to $L_2 \subseteq \Sigma_2^*$ (denoted $L_1 \propto L_2$) if there exists a function that $f: \Sigma_1^* \to \Sigma_2^*$ satisfying:
\begin{itemize}
    \item $f$ is computable by a polynomial DTM time.in polynomial time.
    \item For any $x\in \Sigma_1^*$, $x \in L_1$ if and only if $ f(x) \in L_2 $.
\end{itemize}
\begin{lemma}
If $ L_1 \propto L_2 $, then $ L_2 \in P $ implies $ L_1 \in P $. Alternatively, if $ L_1 \notin P $, then $ L_2 \notin P $.
\end{lemma}
\begin{lemma}
The polynomial reducibility relation is transitive. If $ L_1 \propto L_2 $ and $ L_2 \propto L_3 $, then $ L_1 \propto L_3 $.
\end{lemma}

Polynomial reducibility allows to compare the difficulty of the problem: if $ \Pi_1 \propto \Pi_2 $, then $\Pi_2 $ is at least as hard as an $L_1$. Two problems $\Pi_1$ and $\Pi_2$ are called \emph{polynomially polynomials equivalent} if they $\Pi_1 \propto \Pi_2$ and $\Pi_2 \propto \Pi_1$.

A problem $\Pi$ (or language $L$ is a) \textbf{NP-complete} problem if:
\begin{itemize}
    \item $L\Pi_1 \in NP$ (or $L \in NP$).
    \item Every other language in $\Pi' NP \in NP$, (or language polynomially $ L' \in NP $) reduces is polynomially reducible to $ \Pi $ (or $ L $).
\end{itemize}

NP-complete problems are the hardest in NP. If any NP-complete problem is solvable in polynomial time, then $P = NP $. If $ P \neq NP$, all NP-complete problems lie in $ NP \setminus P $, meaning they have no polynomial-time algorithm algorithms.
\begin{lemma}
If $ L_1 \in NP $ is NP-complete, $ L_2 \in NP $, and $ L_1 \propto L_2 $, then $ L_2 $ is also NP-complete. This simplifies proving NP-completeness proof: by showing that a known NP-complete problem reduces to the a new problem, and that the new problem is in $ \in NP $.
\end{lemma}
\section{Satisfiable Problem}~\label{sec2}
 In this section, we provide classes $P$ and $NP$, polynomial reducibility, and $NP$-complete problems. Theorem on the $NP$-completeness of satisfiability, for Problem $\operatorname{SAT}$. Given a set of Boolean variables $U= \{u_1, u_2, \ldots, u_m\} $ and a set $C$ of clauses over $U$, where each clause is a disjunction of literals (variables $u_i$ or their negations $\neg u_i$).
\begin{question}~\label{question1}
Does there exist a truth assignment $ t: U \to \{T, F\} $ such that all clauses in $ C $ are simultaneously true?
\end{question}
\begin{answer}
We need to prove The Satisfiable problem is NP-complete. Thus, we will discuss following cases. 
\case{1} \textbf{SAT $ \in $ NP}:
    A nondeterministic algorithm for SAT guesses a truth assignment $ t $ for the variables in $ U $ and checks if each clause in $ C $ contains at least one true literal. This verification takes polynomial time (linear in the size of $ C $). Thus, SAT is in $ NP $.
\case{2} \textbf{Every $ L \in NP $ Reduces to SAT}:
    For any language $ L \in NP $, there exists an NDTM $ M $ that recognizes $ L $ in polynomial time $ p(n) $, where $ n $ is the input length. Goal: Construct a polynomial-time computable function $ f_L $ mapping an input $ x \in \Sigma^* $ to a SAT instance (set of variables $ U $ and clauses $ C $) such that $ x \in L $ if and only if $ C $ is satisfiable.\par 
    Construction of $ f_L $, let $ M = (\Gamma, \Sigma, b, Q, q_0, q_Y, q_N, \delta) $ be an $\operatorname{NDTM}$ recognizing $ L $, with $ \Gamma $ as the tape alphabet, $ \Sigma $ as the input alphabet, $ b $ as the blank symbol, $ Q $ as the set of states, $ q_0 $ as the initial state, $ q_Y $ and $ q_N $ as accepting and rejecting states, and $ \delta $ as the transition function. The running time of $ M $ is bounded by $ p(n) $. For an input $ x $ of length $ n $, an accepting computation of $ M $ (if it exists) uses at most $ p(n) $ steps and tape cells numbered from $ -p(n) $ to $ p(n)+1 $.  We define variables:
    \begin{itemize}
        \item $ Q[i, k] $: At time $ i $ ($ 0 \leq i \leq p(n) $), $ M $ is in state $ q_k $ ($ 0 \leq k \leq r $, where $ r = |Q|-1 $).
        \item $ H[i, j] $: At time $ i $, the head scans cell $ j $ ($ -p(n) \leq j \leq p(n)+1 $).
        \item $ S[i, j, k] $: At time $ i $, cell $ j $ contains symbol $ s_k $ ($ 0 \leq k \leq v $, where $ v = |\Gamma|-1 $).
    \end{itemize}
Thus, the set $ C $ is divided into six groups $ G_1 $ to $ G_6 $, each enforcing constraints to ensure a truth assignment corresponds to an accepting computation of $ M $ on $ x $.

    \begin{itemize}
        \item \textbf{Group $ G_1 $}: Ensures exactly one state at time $ i $:
  \begin{equation}\label{eq1}
         \{Q[i, 0], Q[i, 1], \ldots, Q[i, r]\}, \quad 0 \leq i \leq p(n),
  \end{equation}
  Then, when $0 \leq j < j' \leq r$,
\begin{equation}\label{eq2}
\{\overline{Q[i, j]}, \overline{Q[i, j']}\}, \quad 0 \leq i \leq p(n).
\end{equation}
        \item \textbf{Group $ G_2 $}: Ensures exactly one head position at time $ i $ when $0 \leq i \leq p(n)$:
 \begin{equation}\label{eq3}
      \{H[i, -p(n)], H[i, -p(n)+1], \ldots, H[i, p(n)+1]\},
 \end{equation}
 Then, when $-p(n) \leq j < j' \leq p(n)+1$, 
 \begin{equation}
      \{\overline{H[i, j]}, \overline{H[i, j']}\}, \quad 0 \leq i \leq p(n).
 \end{equation}
\item \textbf{Group $ G_3 $}: Ensures exactly one symbol per cell $-p(n) \leq j \leq p(n)+1$:
 \begin{equation}\label{eq5}
     \{S[i, j, 0], S[i, j, 1], \ldots, S[i, j, v]\}, \quad 0 \leq i \leq p(n),
 \end{equation}
 Thus, according to the term $-p(n) \leq j \leq p(n)+1$,
\begin{equation}\label{eq6}
  \{\overline{S[i, j, k]}, \overline{S[i, j, k']}\}, \quad 0 \leq i \leq p(n), \quad  0 \leq k < k' \leq v.
\end{equation}
\item \textbf{Group $ G_4 $}: Sets the initial configuration at time $ i=0 $:
\begin{equation}\label{eq7}
    \{Q[0, 0]\}, \{H[0, 1]\}, \{S[0, j, k_j]\}
\end{equation}
$\text{ for } 1 \leq j \leq n, \text{ where } x = s_{k_1}s_{k_2}\ldots s_{k_n},$ then
\begin{equation}\label{eq8}
    \{S[0, j, 0]\} \text{ for } j < 0 \text{ or } j > n.
\end{equation}
        \item \textbf{Group $ G_5 $}: Ensures the accepting state at time $ p(n) $:
        \[
        \{Q[p(n), 1]\}.
        \]
\item \textbf{Group $ G_6 $}: Ensures valid configuration transitions:
\begin{itemize}
            \item \textbf{Subgroup 1}: If the head is not at cell $ j $, the symbol in $ j $ does not change when $0\leq i \leq p(n)$ and $-p(n) \leq j \leq p(n)+1$:
 \begin{equation}\label{eq9}
      \{\overline{S[i, j, l]}, H[i, j], S[i+1, j, l]\}, \quad 0 \leq l \leq v.
 \end{equation}
            \item \textbf{Subgroup 2:} Transitions follow $ \delta $. For each $ (i, j, k, l) $, where $ 0 \leq i < p(n) $, $ -p(n) \leq j \leq p(n)+1 $, $ 0 \leq k \leq r $, $ 0 \leq l \leq v $, then:
\begin{equation}
\{\overline{H[i, j]}, \overline{Q[i, k]}, \overline{S[i, j, l]}, H[i+1, j+\Delta]\},
\end{equation}
then, 
\begin{equation}\label{eq10}
    \{\overline{H[i, j]}, \overline{Q[i, k]}, \overline{S[i, j, l]}, Q[i+1, k']\},
\end{equation}
Therefore, 
\begin{equation}\label{eq11}
     \{\overline{H[i, j]}, \overline{Q[i, k]}, \overline{S[i, j, l]}, S[i+1, j, l']\},
\end{equation}
\end{itemize}
\end{itemize}
 where $ \delta(q_k, s_l) = (q_{k'}, s_{l'}, \Delta) $ if $ q_k \notin \{q_Y, q_N\} $, else $ \Delta = 0 $, $ k' = k $, $ l' = l $. Then, the correctness of the reduction $f_L(x)$ is established as follows. If $x \in L$, there exists an accepting computation of the nondeterministic Turing machine $M$ on input $x$, which produces a truth assignment that satisfies all clauses in the set $C$. Conversely, if $C$ is satisfiable, there exists a truth assignment that corresponds to an accepting computation of $M$ on $x$, implying that $x \in L$. Therefore, $x \in L$ if and only if $f_L(x)$ is satisfiable.

The polynomial complexity of the reduction $f_L(x)$ is established as follows. The number of variables in the SAT instance, $|U|$, is $O(p(n)^2)$, arising from the variables $Q[i, k]$, $H[i, j]$, and $S[i, j, k]$, which represent the state, head position, and tape contents of the nondeterministic Turing machine at each time step and cell. The number of clauses, $|C|$, is also $O(p(n)^2)$, as the six groups of clauses ($G_1$ to $G_6$) are generated based on the polynomial bound $p(n)$. The encoding length of the SAT instance, defined as Length $[f_L(x)] = |U| \cdot |C|$, is thus $O(p(n)^4)$, which is polynomial in the input size $n$. Constructing $f_L(x)$, including defining the variables and clauses, takes polynomial time because $r$ (the number of states) and $v$ (the size of the tape alphabet) are constants, and $p(n)$ is a polynomial, ensuring the entire reduction process is computationally efficient.

Hence, from~\eqref{eq1}--\eqref{eq11}, $ L \propto L_{\text{SAT}} $ for all $ L \in NP $, we obtained $\operatorname{SAT}$ is NP-complete~\cite{Gary}.  
\end{answer}
\section{Constructing a semigroup with unsolvable equality recognition problem}~\label{sec3}
\begin{question}[Main Question]~\label{question2}
Construct an associative calculus (semigroup) with an algorithmically undecidable word problem, as described in the following theorem:
\begin{theorem}[Post, Markov]~\label{thmn1}
There exists an associative calculus with an algorithmically undecidable word problem.
\end{theorem}
\end{question}
According to Question~\ref{question2} we need to provide some specific question to establish the answer on this question. In an associative calculus $\mathfrak{C}$, the \emph{left divisibility problem} is said to be decidable if there exists an algorithm that, for any two words $\mathfrak{a}, \mathfrak{b}$ in $\mathfrak{C}$, determines whether there exists a solution $\xi$ to the equation $\mathfrak{a} \xi \equiv \mathfrak{b}$ in $\mathfrak{C}$. Similarly, the \emph{left divisibility problem for a fixed element} $\mathfrak{c}$ is said to be decidable in $\mathfrak{C}$ if there exists an algorithm that, for any word $\mathfrak{a}$ in $\mathfrak{C}$, determines whether there exists a solution $\xi$ to the equation $\mathfrak{a} \xi \equiv \mathfrak{c}$ in $\mathfrak{C}$. Show that in the associative calculus $A(\mathfrak{T})$ constructed during the proof of Theorem 3, the left divisibility problem for the element $q_0 v$ is undecidable.    

\begin{answer}
We will start with the associative calculus $A(\mathfrak{T})$ according to Theorem~\ref{thmn1}, where it is constructed based on a Turing machine $\mathfrak{T}$ with symbols $a_0, a_1, \ldots, a_m$ (where $a_0$ represents the blank symbol), states $q_0, q_1, \ldots, q_n$ (with $q_0$ as the initial state), and a set of instructions $\mathfrak{T}$.
Let $\mathcal A=\{a_0,\dots,a_m\}$ be a finite alphabet, whose elements are called in the following \emph{digits}, and let $\mathcal A^*$ be the set of finite words over $\mathcal A$.
The Turing machine $\mathfrak{T}$ is chosen to compute a partial recursive function $E(x)$ that takes values 0 and 1, has no recursive extensions, and whose sets $M_0 = \{ x \mid E(x) = 0 \}$ and $M_1 = \{ x \mid E(x) = 1 \}$ are recursively enumerable but not recursive. Let be define \textbf{alphabet of $A(\mathfrak{T})$:}as the generators are $a_0, a_1, \ldots, a_m, q_0, q_1, \ldots, q_n, v$, where $v$ is an additional symbol.  For each instruction in $\mathfrak{T}$, then for $q_\alpha a_i \rightarrow q_\beta a_j$: $q_\alpha a_i \equiv q_\beta a_j$. For $q_\alpha a_i \rightarrow q_\beta L$: $a_k q_\alpha a_i \equiv q_\beta a_k a_i$ (for all $k = 0, 1, \ldots, m$). Then, for  $q_\alpha a_i \rightarrow q_\beta R$: $q_\alpha a_i \equiv a_i q_\beta$. By considering $a_0 v \equiv v$.

For words $\mathfrak{a}, \mathfrak{b}, \mathfrak{c}, \mathfrak{s}$ in the alphabet $\{a_0, \ldots, a_m\}$, where $\mathfrak{b}$ is empty or ends with a letter other than $a_0$, the relation
\begin{equation}
\mathfrak{a} q_w a_l \mathfrak{b} v \equiv \mathfrak{c} q_0 a_p \mathfrak{s} v
\end{equation}
holds in $A(\mathfrak{T})$ if and only if the Turing machine $\mathfrak{T}$ transitions from the configuration $\mathfrak{a} q_w a_l \mathfrak{b}$ to the configuration $\mathfrak{c} q_0 a_p \mathfrak{s} a_0^x$ (for some $x \geq 0$) in a finite number of steps without extending cells to the left.

Now, we need to prove of Undecidability, the Turing machine $\mathfrak{T}$ computes $E(x)$ such that starting from configuration $q_1 01^x$, it transitions to $q_0 01^{E(x)} 0^s$. Then, for $A(\mathfrak{T})$, we have
\begin{equation}
q_1 01^x v \equiv q_0 01^i v \Leftrightarrow x \in M_i \quad (i = 0, 1).
\end{equation}
Let $\varphi(x)$ be the alphabetic number (a recursive function) of the word $q_1 01^x v$. Define $P_i$ as the set of alphabetic numbers of words equivalent to $q_0 01^i v$. Then,
\begin{equation}
x \in M_i \Leftrightarrow \varphi(x) \in P_i \quad (i = 0, 1).
\end{equation}
Since $M_0$ is not recursive, and $M_0$ is $m$-reducible to $P_0$, the set $P_0$ is not recursive. Similarly, $P_1$ is not recursive.

Now, consider the left divisibility problem for the element $q_0 v$ in $A(\mathfrak{T})$. For a word $\mathfrak{a}$, we need to determine if there exists a solution $\xi$ to the equation $\mathfrak{a} \xi \equiv q_0 v$. Let's consider the specific case where $\mathfrak{a} = q_1 01^x$. Then, we need to determine if there exists $\xi$ such that $q_1 01^x \xi \equiv q_0 v$.

If such a $\xi$ exists, then by the properties of $A(\mathfrak{T})$, the Turing machine $\mathfrak{T}$ must transition from the configuration $q_1 01^x$ to a configuration of the form $q_0 a_0^y$ for some $y \geq 0$. This happens if and only if $x \in M_0$, i.e., $E(x) = 0$.

Therefore, the left divisibility problem for the element $q_0 v$ in $A(\mathfrak{T})$ is equivalent to determining membership in the set $M_0$, which is not recursive. Hence, the left divisibility problem for $q_0 v$ is undecidable~\cite{Maltsev}.
\end{answer}
\section*{Conclusion}
This examination has traversed two fundamental territories in computational complexity theory: the classification of problems through NP-completeness and the exploration of algorithmic undecidability in algebraic structures. Through rigorous mathematical analysis, we have established several significant results that contribute to our understanding of computational boundaries.

The proof of NP-completeness for the Boolean satisfiability problem (SAT) demonstrates the existence of a natural computational problem that encapsulates the difficulty of the entire NP class. By constructing a polynomial-time reduction from any language in NP to SAT, we have reinforced Cook's theorem~\cite{Cook} and illustrated the central role that SAT plays in complexity theory. This reduction technique, further developed by Karp~\cite{Karp} and Levin~\cite{Levin}, provides a powerful framework for classifying computational problems and establishing equivalence classes of difficulty. The detailed construction presented in this work, mapping Turing machine configurations to Boolean formulas, exemplifies the elegant connection between computation and logical satisfiability that lies at the heart of complexity theory.

The investigation of undecidable word problems in semigroups reveals the fundamental limitations of algorithmic approaches to certain algebraic questions. By constructing an associative calculus with an undecidable word problem, following the tradition of Post~\cite{Post} and Markov~\cite{Markov}, we have demonstrated that even in seemingly structured mathematical systems, there exist questions that cannot be resolved by any algorithm. This result connects to the broader landscape of undecidability established by Turing~\cite{Turing} and Church~\cite{Church}, reinforcing the understanding that certain mathematical problems lie beyond the reach of computational methods, regardless of available resources or time.

The juxtaposition of these two areas—NP-completeness and undecidability—provides a comprehensive view of the spectrum of computational difficulty. While NP-complete problems represent challenges that may be solvable with sufficient computational resources (though likely requiring exponential time under the widely believed assumption that P $\neq$ NP), undecidable problems represent absolute barriers to algorithmic solutions. This distinction, as Sipser~\cite{Sipser} notes, helps delineate the boundary between the difficult and the impossible in computational mathematics.

The techniques employed in this work—polynomial-time reductions, encoding of Turing machine computations, and construction of algebraic systems with specific properties—demonstrate the rich interplay between different branches of mathematics and theoretical computer science. These methods not only advance our understanding of specific problems but also contribute to the broader theoretical framework for analyzing computational complexity across diverse domains~\cite{Arora}~\cite{Hopcroft}.

As computational complexity theory continues to evolve, the fundamental questions explored in this work remain central to our understanding of the nature and limits of computation. The P versus NP problem persists as one of the greatest unsolved challenges in mathematics, with implications that extend far beyond theoretical computer science into domains such as cryptography, optimization, and artificial intelligence~\cite{Fortnow}. Similarly, the study of undecidability continues to reveal surprising connections between logic, algebra, and computation, challenging our intuitions about what can and cannot be algorithmically determined.

\end{document}